\documentclass[pra,twocolumn,showpacs,preprintnumbers,amsmath,amssymb]{revtex4-1}
\usepackage{epsfig,amsmath,amssymb,graphicx,amscd,float, overpic}
\usepackage{slashbox}
\newcommand{\pCTR}{\mbox{pCTR}}

\newcommand{\bw}{{\bf w}}
\newcommand{\bc}{{\bf c}}
\newcommand{\bd}{{\bf d}}
\newcommand{\bdd}{{\bf \bar{d}}}
\newcommand{\bK}{{\bf K}}

\newcommand{\pfrac}{\displaystyle\frac}

\newcommand{\secRef}[1]{Sec.\ \ref{#1}}
\newcommand{\tabRef}[1]{Table.\ \ref{#1}}
\newcommand{\tabRefTwo}[2]{Tables.\ \ref{#1} vs \ref{#2}}
\newcommand{\tabRefThree}[3]{Tables.\ \ref{#1}, \ref{#2} and \ref{#3}}

\begin{document}

\author{Xinle Liu}
\email{lxinle@sas.upenn.edu, liuxl@google.com}
\affiliation{Google Inc., 1600 Amphitheatre Parkway, Mountain View, CA 94043}

\begin{abstract}
Selecting the best set of ads is critical for advertisers for a given set of keywords,
which involves the composition of ads from millions of candidates.
While click through rates (CTRs) are important, there could be high correlation among different ads,
therefore the set of ads with top CTRs does not necessarily maximize the number of clicks.
  Greedy algorithm\cite{Cormen:2009:IAT:1614191} has been a standard and straightforward way to find out \textit{a} decent enough solution,
however, it is not guaranteed to be the global optimum.
In fact, it proves not to be \textit{the} global optimum
more than $70\%$ of the time across all our simulations,
implying that it's very likely to be trapped at a local optimum.
In this paper,
we propose a Greedy-Power Algorithm to find out the best set of creatives,
that is starting with the solution from the conventional Greedy Algorithm,
one can perform another Greedy Algorithm search on top of it,
with the option of a few or even infinite rounds.
The Greedy-Power algorithm is guaranteed to be \textit{not} worse,
as it only moves in the direction to increase the goal function.
We show that Greedy-Power Algorithm's performance is \textit{consistently} better,
and reach the conclusion that it \textit{is} able to
perform better than the Greedy Algorithm systematically.
\end{abstract}

\title{How to select the best set of ads: Can we do better than Greedy Algorithm?}
\maketitle

\section{Introduction}
Ads has been contributing to more than $90\%$ of revenue for companies
as Google, Facebook, \textit{etc.},
therefore improving the performance deserves lots of time and efforts,
not only for advertiser themselves, but also for various platform providers as Google.
While advertisers have been spending lots of efforts generating ads with higher quality,
it is never satisfying enough with multiple reasons.
It's subtle to find out the difference of two ads,
especially when the similarity is high,
humans can hardly find out the minor performance difference,
especially without a through understanding of users,
even when they're expressing strong interests in their products.
Also it suffers from the notorious scaling issue,
since ads human writers' efforts do \textit{not} scale up as machines in the modern era of big data.

\section{Methodology}
\subsection{Formulation of the problem}
For each ads group, advertisers could specify a set of keywords to be matched:
\begin{equation}
\label{eq:keywords}
\bw \equiv ( \cdots w_i \cdots), i = 0, 1, \cdots W- 1
\end{equation}

\noindent
and provide a set of creatives:
\begin{equation}
\label{eq:creatives}
\bc \equiv ( \cdots c_j \cdots), c = 0, 1, \cdots N- 1
\end{equation}

With those two as the two dimensions we define the clicks matrix as below:
\begin{eqnarray}
\label{eq:clicks}
\bK \equiv
\left(
  \begin{array}{cccc}
    \cdots & \cdots & \cdots \\
    \cdots & K_{ij} & \cdots \\
    \cdots & \cdots & \cdots
  \end{array}
\right)
\end{eqnarray}

Similarly we could define the $\pCTR$ matrix.

Given the limit on number of creatives for each ads group,
there could only be $M$ creatives to be the final candidates,
i.e. it is to select a subset of $M$ elements from $\bc$:
\begin{equation}
  \bd= (\cdots \bd_{a} \cdots), a = 0, 1, \cdots M-1
\end{equation}
to maximize the number of clicks, with goal function:
\begin{equation}
  G(\bd) = \sum\limits_i \mbox{max}(\{K_{ia}, a = 0 \cdots M-1\})
\end{equation}

The shorthand notation of Greedy Algorithm $G(W, N, M)$
and exact solution of $E(W, N, M)$ will be used through this paper.
The main assumption is that 
for each keyword request from external user,
there could be only one creative to be selected,
and we require it to be \textit{the} one with the largest $\pCTR$,
so as to maximize revenue.

\subsection{Greedy Algorithm}
Greedy Algorithm has been a popular approach for many practical problems,
in some cases it proves to be the optimal solution,
while in most cases it can only provide a \textit{local} optimal solution,
though it is decent enough and
could be a very good approximation to the exact optimal solution.

Greedy Algorithm with size $M$ is similar to mathematical induction,
starting with the same problem with smallest size, normally $0$ or $1$,
one can go one step further for each iteration,
and after $M$ steps it solves the original problem,
though not guaranteed to be \textit{globally} optimal.

Greedy Algorithm for our problem works as the following:
\begin{enumerate}
  \item $a = 1$: select creative $\mbox{argmax}_j \sum\limits_i K_{ij}, j \in [0, N-1]$,
    i.e. the column with the largest sum over all keywords;
  \item $\cdots$
  \item $a = a$: Select the creative $\mbox{argmax}_j G(\bd \cup
      \{c_{j}, j \in [0, N - 1], c_j \notin \bd\}
      )$;
  \item $\cdots$
  \item Until $M$ creatives are selected.
\end{enumerate}

\subsection{Greedy-Power Algorithm: $G^n(r, f; W, N, M)$}
Assuming that Greedy Algorithm is a decent approximation to the exact solution $\bdd$,
it is likely that the difference between those two sets are small,
i.e. the number of creatives in the difference set is $m$, which is much smaller than $M$.
In other words, starting with a solution from Greedy Algorithm,
one only needs to perform a minor tuning and to replace a few creatives if ever necessary,
when greedy algorithm is different from that exact solution $\bdd$.
Here we introduce three parameters,
$r$ as the number of creatives removed from the original solution,
$f$ as the number of creative subsets as starting points,
and $n$ as the number of iterations on top of existing solution.

%% $r$
$r$: With $r$ creatives removed from the original Greedy solution,
one gets a new starting point and
can perform another round of greedy algorithm with problem size $M- r \to M$.
It has the same complexity to the original Greedy Algorithm with a factor of $\pfrac{r}{M} < 1$.
Note that the two solutions can be different by at most $r$ creatives.

%% $f$
$f$: Removing $r = 1, 2, 3 \cdots$ creative has
$M,  M (M-1)/2, M(M-1)(M-2)/6 \cdots$ options respectively.
When $r = 1$, the cost is essentially the same to original Greedy Algorithm run with $M$ iterations,
since here it is $M$ runs with \textit{one} single iteration for each run.
When $r > 1$, there could be a few options:
\begin{itemize}
  \item Systematic approach to enumerate all combinations, with cost increasingly quickly with $r$,
    and much more expensive than the original Greedy Algorithm solution;
  \item By sampling $f$ unique combinations from all available options,
    one can keep the cost comparable to original one.
\end{itemize}
The $r> 1$ option could be better than $r= 1$, since it has the ability to remove multiple elements together,
while the latter option can at most replace $1$ creative.
A simple calculation for a solution with elements $c_x$ and $c_y$ and an even cut,
i.e. $r = M / 2$ and $M$ is even, the probability of $c_x$ and $c_y$ not being in the same half is:
\begin{equation}
  P(\{c_x\}, \{c_y\})= \pfrac{C_{M-2}^{r- 1}}{C_{M-1}^{r-1}} = \pfrac{M-r}{M-1} \sim \pfrac{1}{2}
\end{equation}
At the same time, the probability of $c_x$ and $c_y$ in the same half is also roughly $\pfrac{1}{2}$,
so that they could be simultaneously removed when necessary.
Therefore, a larger $r$ has much more flexibility than a smaller $r$.

%% $n$
$n$: With $r$ creatives replaced by the \textit{2nd} round of Greedy Algorithm,
one can ask the same question again,
can we do better than the current solution?
Actually the question is exactly the same to what we have been asking with a Greedy Algorithm in the first step.
In fact, the new solution is from Greedy Algorithm as well,
and essentially no different from its counterpart in the \textit{first} round.
As long as the goal function keeps increasing,
one can continue this process,
until goal function is the same to the previous iteration,
which implies that we're \textit{not} able to go further,
and it is equivalent to $n = \infty$ in this case.

With those $3$ parameters,
we'd use the notation $G^n(r, f; W, N, M)$ for our proposed Greedy-Power Algorithm.

\section{Simulation}
With the Greedy-Power Algorithm $G^n(r, f; W, N, M)$,
we're able to run some simulations comparing against its baseline Greedy Algorithm.

\subsection{Simulation Setup}
\label{subsec:setup}
\begin{itemize}
  \item Matrix elements are generated from normal distribution and then take their absolute value;
  \item Matrix sizes are at the order of $50 \times 500$ with $M <= 10$;
  \item By default, all simulations are ran multiple times with trajectories
    $T= 500$ for each simulation, repeating $3$ times. % to reduce fluctuations in the final result;
\end{itemize}

\subsection{Simulation Results and Discussion}

\subsubsection{$G^2(r, \cdot; \cdots)$}
\label{subsec:benchmark}

Note that as $r$ increases,
the computational cost increases by a factor of $r$,
with $r$ iterations compared with one single iteration when $r= 1$.
Strictly speaking, we are supposed to compare performance of
$G^n(1, M; W,N,M)$ vs
$G^n(r, M/r; W,N,M)$,
since the computational cost match.
However, the search space size increases exponentially with $r$,
while a factor of $\pfrac{1}{r}$ would effectively remove more options,
resulting in a quickly decreasing coverage ratio in the search space.
Therefore, we'd remove the $\pfrac{1}{r}$ factor
and keep the default value $f= M$ unless otherwise noted,
i.e.\ run same number of creative subsets from the starting solution obtained from the Greedy Algorithm.

\begin{table}[!ht]
\begin{tabular}{|l|l|l|lll}
  \hline
  \# & Matched $(\%)$ & Improvement $(\%)$\\
  \hline
  $1$ & $31.40$ & $1.19$\\
  \hline
  $2$ & $30.20$ & $1.13$\\
  \hline
  $3$ & $28.80$ & $1.14$\\
  \hline
\end{tabular}
\caption{$G^2(1; 30, 300, 6)$ simulation results.}
\label{tab:base:r1}
\end{table}

% I1211 23:24:58.512826 140237409155520 simulation_hybrid.py:119] Simulations #= 500 for matrix (30, 300), M= 6. SquareDiff=20. Matched: 25.60%, clicks ratio= 1.34%.
% I1211 23:27:23.791829 140275908909504 simulation_hybrid.py:119] Simulations #= 500 for matrix (30, 300), M= 6. SquareDiff=20. Matched: 28.60%, clicks ratio= 1.43%.
% I1211 23:28:42.822514 140154569533888 simulation_hybrid.py:119] Simulations #= 500 for matrix (30, 300), M= 6. SquareDiff=20. Matched: 28.80%, clicks ratio= 1.44%.
\begin{table}[!ht]
\begin{tabular}{|l|l|l|lll}
  \hline
  \# & Matched $(\%)$ & Improvement $(\%)$\\
  \hline
  $1$ & $25.60$ & $1.34$\\
  \hline
  $2$ & $28.60$ & $1.43$\\
  \hline
  $3$ & $28.80$ & $1.44$\\
  \hline
\end{tabular}
\caption{$G^2(2; 30, 300, 6)$ simulation results.}
\label{tab:base:r2}
\end{table}

% I1211 23:17:55.788681 140405651945920 simulation_hybrid.py:119] Simulations #= 500 for matrix (30, 300), M= 6. SquareDiff=30. Matched: 26.20%, clicks ratio= 1.71%.
% I1211 23:21:23.531507 140592735803840 simulation_hybrid.py:119] Simulations #= 500 for matrix (30, 300), M= 6. SquareDiff=30. Matched: 24.80%, clicks ratio= 1.61%.
% I1211 23:23:13.360512 140037744736704 simulation_hybrid.py:119] Simulations #= 500 for matrix (30, 300), M= 6. SquareDiff=30. Matched: 23.60%, clicks ratio= 1.51%.
\begin{table}[!ht]
\begin{tabular}{|l|l|l|lll}
  \hline
  \# & Matched $(\%)$ & Improvement $(\%)$\\
  \hline
  $1$ & $26.20$ & $1.71$\\
  \hline
  $2$ & $24.80$ & $1.61$\\
  \hline
  $3$ & $23.60$ & $1.51$\\
  \hline
\end{tabular}
\caption{$G^2(3; 30, 300, 6)$ simulation results.}
\label{tab:base:r3}
\end{table}

From those $G^2(r; 30, 300, 6) \equiv G^2(r, 6; 30, 300, 6)$ simulation data with $r= 1, 2, 3$ in
\tabRefThree{tab:base:r1}{tab:base:r2}{tab:base:r3},
one can clearly see that as $r$ increases,
the likelihood to go out of the local optimum increases,
as the matched ratio decreases.
Also note that increasing $r$ has another benefit,
among those trajectories going out of the optimum,
the overall improvement compared with the benchmark $G(W, N, M)$ also increases.
Therefore, one is confident that the computational cost of
an extra and simple factor $r$ is worthwhile compared with $r= 1$.

\subsubsection{$G^2(\cdot\ $f$; \cdots)$}

% I1212 11:10:28.187001 139755428526528 simulation_hybrid.py:127] Simulations #= 500 for matrix (30, 300), M= 6. SquareDiff=30. Matched: 17.80%, clicks ratio= 1.71%.
% I1212 11:14:48.655678 139945073232320 simulation_hybrid.py:127] Simulations #= 500 for matrix (30, 300), M= 6. SquareDiff=30. Matched: 18.20%, clicks ratio= 1.81%.
% I1212 11:20:05.469158 139701448333760 simulation_hybrid.py:127] Simulations #= 500 for matrix (30, 300), M= 6. SquareDiff=30, factor= 2. Matched: 19.40%, clicks ratio= 1.79%.
\begin{table}[!ht]
\begin{tabular}{|l|l|l|lll}
  \hline
  \# & Matched $(\%)$ & Improvement $(\%)$\\
  \hline
  $1$ & $17.80$ & $1.71$\\
  \hline
  $1$ & $18.20$ & $1.81$\\
  \hline
  $1$ & $19.40$ & $1.79$\\
  \hline
\end{tabular}
\caption{$G^2(3, 12; 30, 300, 6)$ simulation results.}
\label{tab:f2:r3}
\end{table}

% I1211 22:25:44.481442 139937718323648 simulation-hybrid.py:119] Simulations @= 500 for matrix (30, 300), M= 6. SquareDiff=3. Matched: 16.20@@, clicks ratio= 1.89@@.
% I1211 22:33:17.199284 139725773834688 simulation-hybrid.py:119] Simulations @= 500 for matrix (30, 300), M= 6. SquareDiff=3. Matched: 16.20@@, clicks ratio= 1.90@@.
% I1211 22:37:52.531012 140636454357440 simulation-hybrid.py:119] Simulations @= 500 for matrix (30, 300), M= 6. SquareDiff=3. Matched: 15.00@@, clicks ratio= 1.87@@.
\begin{table}[!ht]
\begin{tabular}{|l|l|l|lll}
  \hline
  \# & Matched $(\%)$ & Improvement $(\%)$\\
  \hline
  $1$ & $16.20$ & $1.89$\\
  \hline
  $2$ & $16.20$ & $1.90$\\
  \hline
  $3$ & $15.00$ & $1.87$\\
  \hline
\end{tabular}
\caption{$G^2(3, 18; 30, 300, 6)$ simulation results.}
\label{tab:f3:r3}
\end{table}

Note that $r = 1$ has only $M$ candidates in the search space,
while they're all covered by the default choice of $f= M$,
to match the computational cost of a conventional Greedy Algorithm.

\tabRefThree{tab:base:r3}{tab:f2:r3}{tab:f3:r3} present the extra value
one can get with $f= M, 2M, 3M$ for $r= 3$.
While $f = M \to 2M$ improves the probability
to get of local optimum significantly by close to $10\%$,
the value added on average is not that impressing.
At the same time, $f= 2M \to 3M$ shows less value,
with a minor improvement of both metrics,
which implies the selection of $f$ is more of an art,
as the trade-off between computational cost and added value.

\subsubsection{$G^2(\cdots; \cdots M)$}
% I1211 22:46:53.909145 140366189640128 simulation_hybrid.py:119] Simulations #= 500 for matrix (30, 300), M= 10. SquareDiff=1. Matched: 11.00%, clicks ratio= 0.94%.
% I1211 22:49:16.952120 140281336232384 simulation_hybrid.py:119] Simulations #= 500 for matrix (30, 300), M= 10. SquareDiff=1. Matched: 12.00%, clicks ratio= 0.93%.
% I1211 22:51:17.790704 139779731790272 simulation_hybrid.py:119] Simulations #= 500 for matrix (30, 300), M= 10. SquareDiff=1. Matched: 10.40%, clicks ratio= 0.96%.
\begin{table}[!ht]
\begin{tabular}{|l|l|l|lll}
  \hline
  \# & Matched $(\%)$ & Improvement $(\%)$\\
  \hline
  $1$ & $11.00$ & $0.94$\\
  \hline
  $2$ & $12.00$ & $0.93$\\
  \hline
  $3$ & $10.40$ & $0.96$\\
  \hline
\end{tabular}
\caption{$G^2(1; 30, 300, 10)$ simulation results.}
\label{tab:M10:r1}
\end{table}

% I1211 23:16:09.616672 139736429795776 simulation_hybrid.py:119] Simulations #= 500 for matrix (30, 300), M= 10. SquareDiff=30. Matched: 9.80%, clicks ratio= 1.24%.
% I1211 23:35:20.994626 140673986197952 simulation_hybrid.py:119] Simulations #= 500 for matrix (30, 300), M= 10. SquareDiff=30. Matched: 11.00%, clicks ratio= 1.27%.
% I1211 23:39:34.604046 139628297411008 simulation_hybrid.py:119] Simulations #= 500 for matrix (30, 300), M= 10. SquareDiff=30. Matched: 10.40%, clicks ratio= 1.19%.
\begin{table}[!ht]
\begin{tabular}{|l|l|l|lll}
  \hline
  \# & Matched $(\%)$ & Improvement $(\%)$\\
  \hline
  $1$ & $ 9.80$ & $1.24$\\
  \hline
  $2$ & $11.00$ & $1.27$\\
  \hline
  $3$ & $10.40$ & $1.19$\\
  \hline
\end{tabular}
\caption{$G^2(3; 30, 300, 10)$ simulation results.}
\label{tab:M10:r3}
\end{table}

Comparing with $M= 10$ vs $M= 6$ data as shown in
\secRef{subsec:benchmark}
with the same $r= 1, 3$,
\tabRef{tab:base:r1} vs \tabRef{tab:M10:r1} and
\tabRef{tab:base:r3} vs \tabRef{tab:M10:r3},
it shows that when $M$ increases,
the $r= 1$ option $3$ times less likely to be trapped at local optimum,
which also implies that Greedy Algorithm is very unlikely to be the global optimum $P^{\mbox{\tiny opt}}_{\mbox{\tiny Greedy}}< 10\%$.

Comparing with $r = 1$ vs $r = 3$ results for $M= 10$ in
\tabRefTwo{tab:M10:r1}{tab:M10:r3},
again it shown that a larger $r$ shows extra value to improve
both the probability of getting out of the local optimum and
the ability to find a better optimum based on the simple Greedy Algorithm.

\subsubsection{$G^2(\cdots; W, N \cdots)$}
% I1212 10:00:16.951652 139818493491648 simulation_hybrid.py:119] Simulations #= 500 for matrix (100, 300), M= 6. SquareDiff=1. Matched: 33.00%, clicks ratio= 0.58%.
% I1212 10:05:25.933021 140652637157824 simulation_hybrid.py:119] Simulations #= 500 for matrix (100, 300), M= 6. SquareDiff=1. Matched: 34.00%, clicks ratio= 0.59%.
% I1212 10:09:57.538790 139831651842496 simulation_hybrid.py:119] Simulations #= 500 for matrix (100, 300), M= 6. SquareDiff=1. Matched: 37.80%, clicks ratio= 0.58%.
\begin{table}[!ht]
\begin{tabular}{|l|l|l|lll}
  \hline
  \# & Matched $(\%)$ & Improvement $(\%)$\\
  \hline
  $1$ & $33.00$ & $0.58$\\
  \hline
  $2$ & $34.00$ & $0.59$\\
  \hline
  $3$ & $37.80$ & $0.58$\\
  \hline
\end{tabular}
\caption{$G^2(1; 100, 300, 6)$ simulation results.}
\label{tab:K100:r1}
\end{table}

\tabRefTwo{tab:base:r1}{tab:K100:r1} are different by $3$ times on row size,
i.e. the number of keywords space.
Simulation suggests that both metrics get worse when $K$ increases.
$P_{\mbox{\tiny trapped}}$ increases only a little bit,
implying a Greedy Square Algorithm is slightly more likely to be trapped locally under a larger keywords space.
Also the gain from the Greedy Square Algorithm is getting worse, reduced by a factor of $2$.
This is kind of expected,
since each minimal iteration to add one more creative is exact in the keyword dimension,
while the uncertainty lies mostly if not all in the creatives dimension.
In this case,
one might consider searching for a larger space,
possibilities are to increase $r$, $f$, i.e. the size of the search space.

% I1212 09:40:29.357429 140083751053760 simulation_hybrid.py:119] Simulations #= 500 for matrix (30, 1000), M= 6. SquareDiff=1. Matched: 25.60%, clicks ratio= 1.12%.
% I1212 09:43:22.876621 140615487162816 simulation_hybrid.py:119] Simulations #= 500 for matrix (30, 1000), M= 6. SquareDiff=1. Matched: 25.60%, clicks ratio= 1.25%.
% I1212 09:52:49.660605 140193367165376 simulation_hybrid.py:119] Simulations #= 500 for matrix (30, 1000), M= 6. SquareDiff=1. Matched: 28.00%, clicks ratio= 1.22%.
\begin{table}[!ht]
\begin{tabular}{|l|l|l|lll}
  \hline
  \# & Matched $(\%)$ & Improvement $(\%)$\\
  \hline
  $1$ & $25.60$ & $1.12$\\
  \hline
  $2$ & $25.60$ & $1.25$\\
  \hline
  $3$ & $28.00$ & $1.22$\\
  \hline
\end{tabular}
\caption{$G^2(1; 30, 1000, 6)$ simulation results.}
\label{tab:N1000:r1}
\end{table}

\tabRefTwo{tab:base:r1}{tab:N1000:r1} are different by $3$ times on column size,
i.e. the number of creatives space.
Simulation suggests that both metrics improve when $N$ increases,
an indication that a Greedy Square Algorithm is more likely to improve under a larger creatives space.

\subsubsection{$G^n(\cdots; \cdots)$}
Given the solution $g^1$ from conventional Greedy Algorithm,
we should be able to run another round of Greedy Algorithm on top it,
resulting in another solution $g^2$ from the Greedy Square Algorithm.
Given the fact that $g^2$ is a solution resulting from the Greedy Algorithm,
there's essentially little or even no difference vs $g^1$.
That's, starting with the Greedy Algorithm solution in iteration $i$, $g^i$,
one can always run another round of Greedy Algorithm to obtain a solution $g^{i+1}$.

There might be two outcomes:
\begin{itemize}
  \item $g^{i+ 1}$ is worse than $g^i$, keep $g^i$ and stop here;
  \item $g^{i+ 1}$ is equivalent to (starting in a new direction \textit{could} find something new)
    or better than $g^i$, keep $g^{i+ 1}$;
\end{itemize}

Note that when the stopping rule of
$g^{i+ 1}$ being worse than $g^i$ is triggered,
$G^n$ is equivalent to $G^{\infty}$,
which would be equivalent to the exact solution ideally,
though it \textit{could} be trapped somewhere in a local optimum in principle.

\section{Conclusions}
We present a Generalized Greedy Algorithm,
i.e.\ the Greedy-Power Algorithm $G^n(r, f; W,N,M)$,
characterized by three parameters:
number of creatives to remove $r$ from a given solution implying a difference of up to $r$ creatives from original solution,
number of branches to take for a given solution $f \sim M$,
and number of iterations for improvement $n$ implying a Greedy Algorithm with a power of $n$.
With $f \sim M$,
we effectively impose that any following improvement step
should have the same complexity of a conventional Greedy Algorithm,
up to a factor of $r$,
and achieve solutions consistently better than the latter.
% From the probabilistic perspective,
% we gives a theoretical explanation of why Greedy Power Algorithm should be able to work.
As such, with twice (and an additional factor of $r$) the cost of a Greedy Algorithm,
one is likely to achieve decent added value from the Greedy-Power Algorithm,
especially when the dimension $M$ is large,
when a Greedy Algorithm is more likely to be trapped at a local optimum,
and corrected by the Greedy-Power Algorithm.

\subsection{Future Directions}
The setup of the Greedy-Power Algorithm $G^n(r, f; W, N, M)$ is all done,
while the effectiveness with real data is still under exploration,
and it'd be interesting to see any significant difference of improvement
from the random standard normal distribution.
Also, it would provide more insights on the performance of Greedy Algorithm,
and how it walks to the global optimum with a big $n$,
which would be equivalent to the $n\to \infty$ when there is no more gain from the goal function.

At the same time,
there could be other approaches for this problem setup,
and one alternative is to find the exact solution is completely from a matrix perspective,
which will be discussed in a separate paper in the near future.

\section{References}
%% \bibliography{greedy_square}

\end{document}